\documentclass{hotnets2020}

\usepackage{times}
\usepackage{color}
\usepackage{hyperref}
\usepackage{subfigure}
\usepackage{url}
\usepackage{cleveref}
 \allowdisplaybreaks
 \usepackage{xspace}

\usepackage{booktabs}
\usepackage{colortbl}
\usepackage{float}
\usepackage{listings}
\usepackage{mathtools}

\usepackage{listings}
\newcounter{insightlabel}
\newcounter{insightnmbr}
\renewcommand{\theinsightlabel}{\em \textbf{\theinsightnmbr}}
\newenvironment{insight}{
\begin{list}{\textbf{\em Problem }\theinsightlabel:~}{\usecounter{insightlabel}\stepcounter{insightnmbr}\setlength{\labelwidth}{0pt}\setlength{\labelsep}{0pt}\setlength{\leftmargin}{0in}\noindent\rule{\linewidth}{1pt}\vspace{-9pt}\item \em}}{\\[-7pt]\end{list}\vspace{-11pt}\noindent\rule{\linewidth}{1pt}}

\newenvironment{mycorollary}
{\begin{list}{\textbf{\em Extension:}~}{\setlength{\labelwidth}{0pt}\setlength{\labelsep}{0pt}\setlength{\leftmargin}{0in}\noindent\vspace{-10pt}\item \em}}{\\[-7pt]\end{list}\vspace{2pt}
}

\lstdefinelanguage{GAP}{
 keywords={Select, Distinct,From,HeavyHitter,Where},
 keywordstyle=\color{black}\bfseries,
 morekeywords={[2]Subgraph,Triangle,FourClique,isAdjacent},
 keywordstyle={[2]\color{black}\bfseries},
 morekeywords={[3]if,return,else,None,False,True},
 keywordstyle={[3]\bfseries},
 basicstyle=\footnotesize\ttfamily,
 identifierstyle=\color{black},
 sensitive=false,
 comment=[l]{\/\/},
 morecomment=[s]{/*}{*/},
 commentstyle=\color{black}\ttfamily,
 stringstyle=\color{red}\ttfamily,
 breaklines=true,
}

\lstnewenvironment{codefragment}
{\lstset{
    language=GAP,
    frame=top,
    numbers=left,
    columns=fullflexible,
    postbreak=\mbox{\textcolor{red}{$\hookrightarrow$}\space},
    captionpos=b}}
    {}
 \lstnewenvironment{codefragment2}
{\lstset{
    language=GAP,
    frame=bottom,
    columns=fullflexible,
    postbreak=\mbox{\textcolor{red}{$\hookrightarrow$}\space},
    captionpos=b}}
 {}

\usepackage{algorithm}
\usepackage[noend]{algpseudocode}

\newenvironment{packeditemize}{
\begin{list}{$\bullet$}{
\setlength{\itemsep}{1.5pt}
\setlength{\labelwidth}{8pt}
\setlength{\leftmargin}{10pt}
\setlength{\labelsep}{3pt}
\setlength{\listparindent}{\parindent}
\setlength{\parsep}{1.5pt}
\setlength{\parskip}{1.5pt}
\setlength{\topsep}{1.5pt}}}{\end{list}}

\newcommand{\NO}{NO\xspace}
\newcommand{\AD}{AD\xspace}
\newcommand{\PD}{PVD\xspace}

\newtheorem{innercustomgeneric}{\customgenericname}
\providecommand{\customgenericname}{}
\newcommand{\newcustomtheorem}[2]{%
  \newenvironment{#1}[1]
  {%
   \renewcommand\customgenericname{#2}%
   \renewcommand\theinnercustomgeneric{##1}%
   \innercustomgeneric
  }
  {\endinnercustomgeneric}
}

\newcustomtheorem{corollary}{Corollary}

\newcommand{\tightcaption}[1]{\vspace{-0.2cm}\caption{#1}\vspace{-0.2cm}}

\newcommand{\para}[1]{\smallskip\noindent\textbf{#1}}
\newcommand{\paraf}[1]{\noindent\textbf{#1}}
\newcommand{\cut}[1]{}

\definecolor{orange}{RGB}{255, 69, 0}

\hypersetup{pdfstartview=FitH,pdfpagelayout=SinglePage}

\begin{document}


\title{Sketchy With a Chance of Adoption:  \\ 
 Can Sketch-Based Telemetry  Be Ready for Prime Time?}

\author{Zaoxing Liu$^{1,2}$, Hun Namkung$^{1}$, Anup Agarwal$^{1}$, Antonis Manousis$^{1}$,\\ Peter Steenkiste$^{1}$, Srinivasan Seshan$^{1}$, Vyas Sekar$^{1}$\\
\large\textit{$^{1}$Carnegie Mellon University, $^{2}$Boston University}}
\maketitle

\begin{abstract}

Sketching algorithms or sketches have emerged as a promising alternative to the traditional packet sampling-based network telemetry solutions. At a high level, they are attractive because of their high resource efficiency and accuracy guarantees. While there have been significant recent advances in various aspects of sketching for networking tasks,  many fundamental challenges remain unsolved that are likely stumbling blocks for adoption. Our contribution in this paper is in identifying and formulating these research challenges across the ecosystem encompassing network operators, platform vendors/developers, and algorithm designers. We hope that these serve as a necessary fillip for the community to enable the broader adoption of sketch-based telemetry.

\end{abstract}
\section{Introduction}
At the core of managing networks, network telemetry plays a crucial role in understanding
what is happening in the network and informing management decisions. For example, to 
improve cloud security, telemetry  enables operators to detect network anomalies and
attacks in a timely fashion. Similarly, in order to optimize traffic engineering and 
ensure that service-level agreements (SLAs) for applications are met, operators commonly
rely on telemetry to monitor network flow distributions. 
Traditionally, flow-based telemetry is done via offline analysis or some form of packet
or flow sampling (e.g., NetFlow~\cite{netflow} and sFlow~\cite{sflow}). However, given
the need
for timely results using constrained compute/memory resources, offline analysis is not a
practical option. Moreover, sampling only provides coarse-grained flow size distributions, and cannot provide accurate results for more fine-grained
key telemetry tasks such as entropy estimation, distinct count, and change
detection~\cite{duffield2003estimating,new_directions,k-ary}.

To address the drawbacks of sampling approaches, sketching algorithms (or sketches for short) have been extensively studied in recent
years (e.g., \cite{bar2002counting,RHHH,CountSketch,CMSketch,SketchVisor,sketchlearn,zhou2019generalized,ivkin2019know,univmon,elasticsketch,HashPipe,liu2019nitrosketch,OpenSketch,Entropy1,k-ary,lean,revsketch}). In light of increasing network traffic and ever-evolving application dynamics, sketches have emerged as a promising solution for real-time network telemetry.

This paper is a reflection  on the current state of sketch-based telemetry  to examine not just what sketch-based systems {\em can}
do but what {\em should} be done to enable broader adoption. To this end, we look
at the state of the sketch-based telemetry ecosystem from the perspective of three
key stakeholders in Figure~\ref{fig:intro}: (1) {\em Network Operators (\NO)}
who are the users/consumers of telemetry capabilities; (2) {\em Algorithm Designers
(\AD)} who design and analyze sketching algorithms; and (3) {\em Platform Vendors and
Developers (\PD)} who provide  hardware/software primitives and APIs in various
platforms (e.g., Intel DPDK~\cite{dpdk}, Barefoot Tofino~\cite{tofino}, Broadcom Trident~\cite{trident}, Mellanox~\cite{mellanox}, among others)
and use these APIs to implement sketch-based functions.

\begin{figure}[t]
\centering
\includegraphics[width=0.88\linewidth]{figures/intro.pdf}
\vspace{-2mm}
\tightcaption{Overview of the problems from the stakeholders in sketch-based telemetry.}
\vspace{-2mm}
\label{fig:intro}
\end{figure}
By taking this ecosystem-level view, we identify four areas of gaps between stakeholder and interaction requirements and existing research (blue boxes in Figure~\ref{fig:intro}): 
    
\begin{packeditemize}
\item {\bf NO-Centric:}  Most existing  efforts assume operators have extensive knowledge about the algorithms and their underlying data structures, which is not realistic.
  There are few, if any, efforts to help operators translate high-level intents into sketches.  This requires both high-level interfaces as 
  well as precise resource management. While NO's intents may involve different sketches and devices, current solutions (e.g., ~\cite{univmon,elasticsketch,sketchlearn,moshref2014dream}) do not consider the composition of multiple types of sketches and the heterogeneity of network devices. 

\item {\bf Between \NO/\AD:}
Prior theoretical work in sketching algorithms covers many common telemetry tasks, and more recent work on general sketches can cover a broad portfolio of tasks~\cite{univmon}. Despite these advances, 
many common \NO intents fall outside the scope of the  literature. For instance, for attack detection, operators are interested in obtaining statistics from not only one dimension of data (e.g., SrcIP) but multiple dimensions (e.g., any subset of the combinations in 5-tuple). Conversely, we find that the theory community has many rich capabilities and streaming models (e.g., turnstile~\cite{li2014turnstile,turnstile1},  sliding-window~\cite{exponential_histogram,braverman2007smooth,zero_one_sliding},  and distributed functional monitoring~\cite{cormode2013continuous, functional_monitor}) that are yet to find practical adoption in networking.

\item {\bf Between \AD/\PD:}  While sketching algorithms are
theoretically lightweight, existing  algorithms  may not  be
efficiently  realizable across diverse platforms as highlighted
by  recent efforts~\cite{SketchVisor,liu2019nitrosketch,elasticsketch,HashPipe,yang2020joltik}. 
Similarly,  while existing languages and APIs~\cite{p4_studio,p4,netronome}
are sufficiently expressive to specify  different sketch algorithms, na\"ive implementations are often resource
intensive, thus  nullifying any potential benefits~\cite{SketchVisor,liu2019nitrosketch}. This suggests
the need for new sketch-centric APIs, language support, and
best practices.

\item {\bf Between \NO/\PD:} Given that the success of the operator's
policies depends crucially on how accurately telemetry reflects current
network conditions, verifying the practical accuracy and correctness 
of sketches at post-deployment is a major priority for the \NO.
In addition, while platform vendors have designed and delivered trusted hardware capabilities 
(e.g., Intel SGX~\cite{intel-sgx}, AMD SEV~\cite{amd-sev}, and ARM TrustZone~\cite{arm-trustzone}) to ensure the integrity
of the program running on the device, the integrity of sketch-based telemetry logic has yet
to be protected.

\end{packeditemize}

Our contribution in this paper is to identify and
formulate challenges that need to be addressed to enable  sketch-based  telemetry to  be more widely adopted. While this list of challenges is by no means 
exhaustive, our goal is  to start the conversation regarding the
ecosystem's missing pieces. We hope that our work will inspire the 
community to tackle these as-yet-unsolved issues, eventually enabling
the practical adoption of sketch-based telemetry.

\section{Background}
In this section, we first provide some background on sketches 
and their use  in network telemetry. We then introduce the key stakeholders in sketch-based telemetry to set the context for the  research challenges.

\subsection{Sketching Algorithms}

Sketching algorithms (sketches) can process data streams accurately and efficiently in an
online fashion. Sketches are attractive for network monitoring precisely because they typically
require small memory footprints to estimate traffic statistics with provable accuracy
guarantees. In addition to  network 
telemetry~\cite{CountSketch,CMSketch,SpaceSavings,Entropy1,univmon,OpenSketch,elasticsketch,SketchVisor,sketchlearn}, sketch-based approaches have also been applied in databases~\cite{mergeable,sigmod2019},
streaming analytics~\cite{apache-druid}, and machine learning~\cite{recursiveDL,ivkin2019communication,jiang2018sketchml}.

Sketches draw on rich theoretical foundations starting from the foundational  ``AMS'' paper~\cite{ams}. At a  high level, the problem they address is as follows: Given an
input stream of {\em <key, value>} pairs (e.g., <5-tuple, packet size> pairs in network traffic), a sketching algorithm is allowed to make a single pass over the data stream
to compute statistics while using sub-linear (usually poly-logarithmic) memory space 
compared to the total size of the dataset and the number of distinct keys. When processing 
each  item in the stream, a sketch  typically maintains a table of counters in the main memory and computes multiple independent hashes to update a small 
random set of counters in the table. These algorithms are backed by rigorous theoretical 
analysis on bounded accuracy-memory tradeoffs for arbitrary workload patterns. 

\para{Sketch-based network telemetry.} Sketches are useful approaches for key network
telemetry  tasks, such as (1) Heavy-Hitter detection to discover large  
flows~\cite{CMSketch,CountSketch,SpaceSavings,HashPipe,univmon,elasticsketch}; (2) Entropy Estimation to analyze traffic distributions for
anomaly detection~\cite{simple_entropy,univmon,Entropy1}; (3) Change Detection to identify significant traffic shifts over time~\cite{k-ary,OpenSketch,univmon}; (4)  Cardinality Estimates to detect the number of distinct items/flows in the network
traffic~\cite{bar2002counting,HLL,univmon,SketchVisor}; (5)  Performance Monitoring to
identify flows with high packet loss, large latency, and high out-of-order or retransmitted packets~\cite{lean}; (6) Superspreader Detection to identify sources that contact many different destinations~\cite{OpenSketch}, among others.

\subsection{Stakeholders for Telemetry Deployment} 
We identify three key players in the ecosystem that drive and influence the adoption
of the above sketch-based telemetry.

\paraf{Network operator:} Network operators rely on 
real time telemetry to make timely decisions 
that ensure network reliability, performance, and security.
To this end, they may want  network-wide information such
as global heavy hitter flows, distinct flows, and entropy
changes on various traffic distributions. 
Ideally, network operators want to express high-level
telemetry objectives without having to worry about  low-level algorithm and implementation  details about sketches. 
 
 \begin{figure}[h]
\centering
    \begin{minipage}[t]{.49\textwidth}
\textbf{Q1: Return 5-tuple 0.005-heavy hitters from all flows}
\begin{codefragment2}
FlowKey = (SrcIP,SrcPort,DstIP,DstPort,Proto)
C="Select HeavyHitter(p.FlowKey,0.05) From *"
return C
  \end{codefragment2}
  \end{minipage}
  \begin{minipage}[t]{.49\textwidth}
  \textbf{Q2: Return distinct DstIP count that a host connects to}
  \begin{codefragment2}
C="Select Distinct(p.dstIP) From *
Where p.srcIP=1.2.3.4"
return C
  \end{codefragment2}
  \end{minipage}
 \vspace{-4mm}
\tightcaption{Examples of envisioned telemetry queries.}
\label{fig:query_example}
\end{figure}
 
 For example, operators may specify queries like Q1 and Q2 depicted in Figure~\ref{fig:query_example}. A telemetry system should provide an  interface  to write queries,   identify if the queries can be  supported by existing primitives, and distribute the monitoring responsibilities efficiently across a network.   If better or new sketches are needed, the telemetry system must pass these information to algorithm designers described below.

\para{Algorithm designers:}  We envision an active community of algorithm designers developing new sketching algorithms to estimate different telemetry metrics. They would like to understand the requirements of network operators to design improved or new sketching algorithms for needed metrics. 
In practice, however, it requires significant efforts to translate theoretical 
algorithms into optimized implementations on diverse platforms. As richer primitives
need to be designed and new platforms (e.g., Barefoot Tofino~\cite{tofino} and Multi-engine
SoC SmartNIC~\cite{netronome}) emerge, algorithm designers increasingly find themselves in need
of a mature sketch-based framework allowing them to develop and evaluate algorithmic tools along with platform vendors/developers described below.

\para{Platform vendors and developers:} Platform vendors offer 
specialized capabilities that implement and optimize sketches on
various hardware and software platforms. For instance, we have already
seen programmable switches, SmartNIC, FPGA, and software switches
established in today's networks, and we envision future deployments
with richer and more diverse platform capabilities. 
Ideally, platform vendors should provide primitives
for these developers to optimally support sketch-based telemetry. However, recent work suggests it is non-trivial 
 to efficiently implement sketches~\cite{liu2019nitrosketch,HashPipe}. In this respect,
we envision the need for these two stakeholders to 
jointly contribute their domain  expertise to achieve optimized sketch
implementations.

\newcommand{\Q}{\ensuremath{\mathcal{Q}}}
\newcommand{\q}{\ensuremath{{q}}}
\newcommand{\Req}{\ensuremath{\mathcal{R_A}}}
\newcommand{\req}{\ensuremath{{r_a}}}
\newcommand{\Perf}{\ensuremath{\mathcal{R_P}}}
\newcommand{\perf}{{r_p}}
\newcommand{\T}{\ensuremath{\mathcal{T}}}
\newcommand{\W}{\ensuremath{\mathcal{W_r}}}
\newcommand{\D}{\ensuremath{\mathcal{D}}}
\newcommand{\Sk}{\ensuremath{{S}}}
\newcommand{\rsd}{\ensuremath{{r_{s,d}}}}
\newcommand{\lsd}{\ensuremath{{l_{s,d}}}}
\newcommand{\csd}{\ensuremath{{c_{s,d}}}}
\newcommand{\cd}{\ensuremath{{c_{d}}}}
\newcommand{\rd}{\ensuremath{{r_{d}}}}
\newcommand{\ld}{\ensuremath{{l_{d}}}}

\section{Research Challenges}
Next, we formulate a broad (but non-exhaustive) list of
open research problems P1$\dots $P6 and some of their extensions for a sketch-based telemetry ecosystem. As depicted
in Figure~\ref{fig:problems}, we conceptually cluster these challenges according
to each stakeholder's needs and considerations.

\begin{figure}[t]
\vspace{-3mm}
\centering
\includegraphics[width=0.66\linewidth]{figures/problems.pdf}
\vspace{-4mm}
\tightcaption{Open problems between stakeholders.}
\vspace{-2mm}
\label{fig:problems}
\end{figure}

\medskip \noindent {\bf Preliminaries:} We introduce some terms and notations to formulate the  problems (summarized in Table~\ref{tab:definition}). 
\begin{packeditemize}
\item The constants represent the inputs to the telemetry system. Specifically, network operators can define their telemetry needs by a list of input constants: (1) Queries $\Q$ consisting of a set of $k$ (potentially infinite) query definitions $\{q_1,\dots,q_k\}$; (2) Requirements $\Req=\{\req^1,\dots, \req^k\}$ defining a set of accuracy requirements (e.g., accuracy target 95\% with 0.99 confidence) for queries $\{\q_1,\dots,\q_k\}$ and similarly $\Perf=\{\perf^1, \dots, \perf^k\}$ as the packet rate requirements; (3) Network characteristics including topology information $\T$, device information with resource capabilities $\D$, and traffic workload characteristics $\W$.
\item The variables are the notations for the intermediate or final outputs of the telemetry system: (1) $\Sk$ is a set of sketch definitions with appropriate memory and flow-key/OD-pair configurations (e.g., a Count-Min sketch tracking 5-tuple flows with 5 x 2048 32-bit counters); (2) $\rsd$ is
the resource configuration of sketch instance $s$ on device $d$ (e.g., assigning 200KB and 2 
cores for $s$ on CPU) and $\lsd$ is the processing latency of $s$ on $d$ (e.g., $1\mu s$ on CPU);
(3) $\csd$ is the implementation (binary code) of sketch instance $s$ on device $d$. When there are multiple sketch instances in $d$, $\cd$ represents the implementation of all
instances combined;
(4) $\rd$ is the actual resource usage of $\cd$ and $\ld$ is the actual processing latency of $\cd$.
\end{packeditemize}

\begin{table}[t]
\centering
\footnotesize

\begin{tabular}{ cl }
\toprule
\textbf{Constants} & Definition\\
\midrule
$\Q$ & Set of telemetry queries\\
$\Req$ & Set of accuracy requirements, \\
& ~~e.g., accuracy target and confidence level\\
$\Perf$ & Set of performance requirements, e.g., packet rate\\
$\T$ & Topology information, e.g., links and devices\\
$\D$ & Set of device instances with resource constraints, \\
& ~~e.g., SmartNIC w/ 4 engines and 10MB SRAM \\
$\W$ & Traffic workload characteristics, e.g., distribution \\

\end{tabular}
\begin{tabular}{ cl }
\toprule
\textbf{Variables} & Definition\\
\midrule
${S}$ & Set of sketch definitions with configurations \\ 
$r_{s,d}$ & Resource config. for sketch s on device $d$  \\
$l_{s,d}$ & Processing latency for sketch s on device $d$ \\
$c_{s,d}$ & Implementation of sketch $s$ on device $d$\\
$c_{d}$ & Implementation of all sketches on device $d$\\
$r_{d}$ & Actual resource usage of device $d$ from $c_{d}$ \\
$l_{d}$ & Actual processing latency of device $d$ from $c_{d}$ 
\\

\bottomrule
\end{tabular}
\vspace{-2mm}
\caption{Summary of notations in problem definitions.}
\vspace{-3mm}
\label{tab:definition}
\end{table}

\subsection{Network Operator-Centric}
\begin{insight}[Query Language]
Is there a high-level declarative language that can precisely define sketch-based telemetry queries
$\Q$?
\end{insight}

Traditionally, sketch-based telemetry is   designed under a narrow scope in  the queries it  supports. Specifically, 
existing frameworks are either designed to support one type of queries~\cite{sigmod2019}
or assume that the operators determine at query time the appropriate (available) 
sketch for each query. For example, to detect Superspreaders  (i.e., SrcIPs that connect
to many distinct DstIPs), the operators need to make a choice between Count-Min + HLL and CountSketch + UnivMon whereas
to conduct change detection they need to choose between K-ary and Count-Min. 
As a result, developing a unified front-end for such telemetry systems was, to the
best of our knowledge, never seen as a key design priority. Specifically, the operators 
should be able to conceptually describe the characteristics of a query to execute (e.g., type of metrics, appropriate aggregation of data, accuracy constraints) without explicitly specifying the execution mechanism.

Existing efforts have proposed several query languages for network telemetry~\cite{sonata,NetQRE,marple}, streaming database~\cite{TelegraphCQ,gigascope}, and traffic analysis~\cite{chimera}. These efforts are self-contained for their systems but may not be an ideal fit for sketch-based telemetry. Specifically, they did not consider sketches as their primitives and overly complicate the query definitions for sketches.
For instance, Sonata~\cite{sonata} can specify the detailed packet-level queries with dataflow operators (e.g., map, filter, reduce) but it is unclear how to describe sketches, and NetQRE~\cite{NetQRE} extends from quantitative regular expressions~\cite{qre} to define flow-level and application-level statistics and polices. In addition, the telemetry tool Marple~\cite{marple} is designed to support a particular set of performance metrics only. Similarly, streaming databases such as Gigascope~\cite{gigascope} support continuous queries over packet headers or counts via a SQL-alike language but do not support other metrics such as network performance and traffic patterns. 

\begin{insight}[Resource Optimization]
Given a set of queries $\Q$ with
accuracy requirements $\Req$ and performance requirements $\Perf$, traffic workload characteristics $\W$, topology $\T$, and device instances $\D$,
generate resource configuration $\rsd \ \forall s,d$ within a time
budget such that $\sum_s\sum_d \rsd$ is minimized and $\forall s \in {S}$
meets $\Req$ and $\Perf$
\end{insight}

Given a set of queries $\Q$, each with associated accuracy and performance 
requirements,  traffic workload characteristics and a network topology, the
operator's high-level goal is to  deploy 
appropriate sketches across the deployment such that SLAs are met while
minimizing overall resource usage. The operator
ideally wants to view their deployment under as ``one-big-switch'' 
 without worrying about   manually distributing
sketches across the various devices in the deployment to ensure appropriate
correctness and coverage.  However, realizing this conceptual goal, requires
addressing a number of sub-challenges which we introduce now and discuss in more
detail in the following subsections:

\begin{packeditemize}
\item {\bf Problem 3:} Translate each  $q \in \Q$ to appropriate sketch
definitions with conservative (traffic-oblivious) memory configurations $\Sk$
to meet accuracy requirements $\Req$. %

\item {\bf Problem 4:} Given a heterogeneous network deployment,
develop optimal device-specific sketch implementations, given 
sketch definitions and configurations $s \in \Sk$. 

\item {\bf Problem 5:} Given traffic workload characteristics $\W$,
optimize each sketch's memory configuration to provide better 
memory-accuracy tradeoff and further reduce resource usage.

\item {\bf Problem 6:} Once sketches are deployed on device $d$,
verify their correctness to ensure the expected accuracy requirements
$\Req$ are met.
\end{packeditemize}

\begin{figure}[t]
\centering
\includegraphics[width=0.66\linewidth]{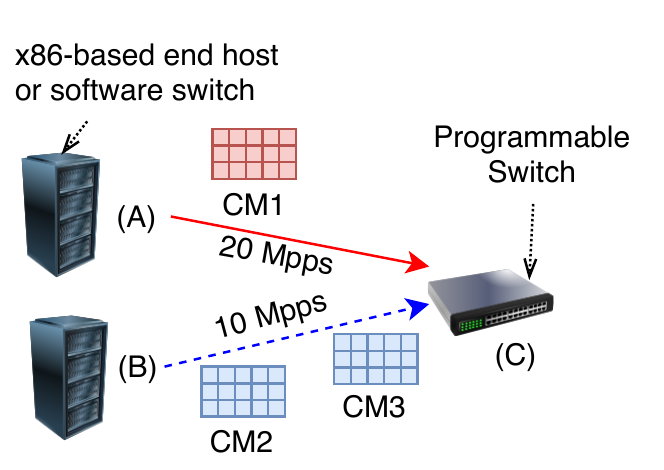}
\includegraphics[width=0.88\linewidth]{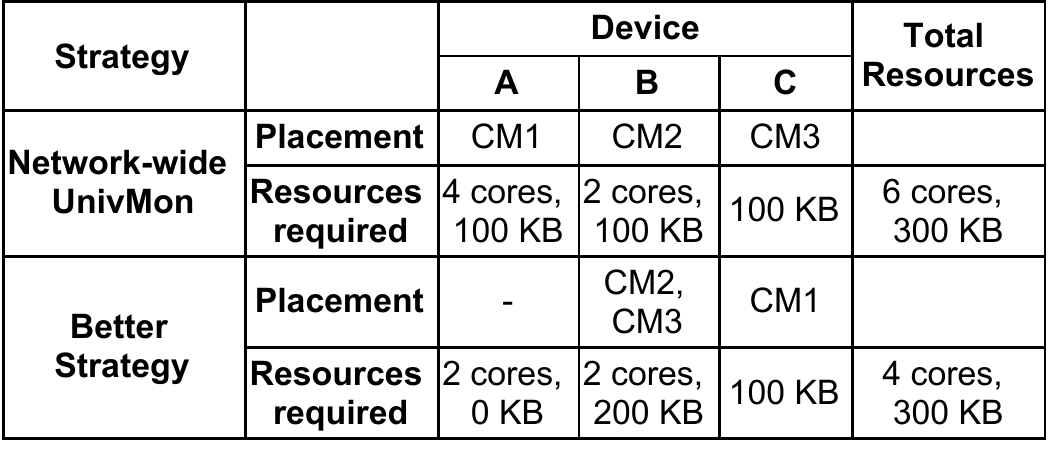}
\vspace{-2mm}
\tightcaption{Example of network-wide UnivMon not optimally placing the sketches.}
\vspace{-1mm}
\label{fig:univmon}
\end{figure}

While prior work presented an early version of a network-wide solution~\cite{univmon}, it  does not take traffic workload characteristics, different types of sketches, and the heterogeneity of the devices into account, and can converge to a sub-optimal or even infeasible sketch placement and resource allocation. Figure~\ref{fig:univmon} shows a simple scenario where network-wide UnivMon does not optimally place three Count-Min sketch instances in a topology of three programmable devices. Specifically, in the example, the operator wants to know the 5-tuple heavy hitters over traffic between devices A and C (CM1) and the heavy hitters over traffic between devices B and C separately for (SrcIP, SrcPort) and (DstIP, DstPort) flow keys (CM2 and CM3). Resource optimization approaches will decide which sketch will be placed on which device while being aware of the resources required for these sketches given different performance requirements for different devices: (1) UnivMon, which is unaware of the interaction between performance requirements and resource usage, tries to balance memory usage by placing a sketch on each device. This results in placing a sketch on device $A$ which sees 20Mpps traffic. In order to accommodate a sketch and support this forwarding rate, device $A$ requires 4 cores. (2) A better strategy shifts telemetry load towards device $B$\footnote{Device B runs in a CPU polling mode.} which sees less traffic and can accommodate 2 sketches while meeting the 10Mpps requirement. Device $A$ in this strategy does not maintain a sketch and only needs 2 cores to maintain 20 Mpps traffic forwarding. Note: Device $C$'s compute resources are the same in both strategies and hence are not shown. 

\begin{mycorollary}[Maximum Performance] Given a set of queries $\Q$ with requirements $\Req$, topology information $\T$, and devices $\D$,  output resource configuration $\rsd$ for all $s,d$ such that $\sum_s\sum_d \lsd$ is minimized and $\forall s \in \Sk$ meets $\Req$
\end{mycorollary}

This extension aims at providing optimized network-wide sketch placement and resource allocation that meets the device resource constraints and minimizes total packet processing overhead. In this optimization, we aim at deploying a telemetry solution to handle the largest possible volume of traffic for given queries, which potentially offers us the ability to monitor bursty traffic. Meanwhile, this type of optimization is useful for operators to control the maximum volume of traffic that goes into the telemetry infrastructure.

\subsection{Network Operator \& Algorithm Designer}
\begin{insight}[Queries to Sketch Definitions] Design
a compiler that translates queries $\Q$ into sketch 
definitions and configurations $\Sk$ that meet accuracy requirements $\Req$ 
\end{insight}

Here, our focus is on translating telemetry queries into a set of
\textit{practical} sketch definitions with memory configurations
satisfying the accuracy requirements from the queries, irrespective
of traffic workload characteristics and hardware platforms. This
is possible because the accuracy guarantees of sketches are hardware 
agnostic and only depend on the memory configuration. Thus, one can
potentially leverage the theoretical analysis from algorithm designers
to provide traffic-oblivious sketch memory configurations. For example, if a query 
specifies a heavy hitter task with $98\%$ accuracy and 0.99 confidence
level, we envision a compiler to generate a platform-agnostic 
sketch configuration (e.g., Count-Min Sketch with $r\times d$ counters)
that maintains errors $\le 2\%$ with 0.99 probability under any workload 
distribution. 

This is the first step towards network-wide
device-aware resource management, which requires target-agnostic memory configurations treating the network-wide 
topology as a ``one-big-switch'' and corresponding
performance characteristics on each hardware target as input.

\begin{mycorollary}
[Expressiveness] If the network operator's telemetry queries $\Q$ cannot be compiled to $\Sk$, can algorithm designers develop new sketching algorithms to address the failures? \end{mycorollary}

While there have been significant advances in developing
sketches for various telemetry tasks, the intents of network 
operators may still fall outside those of existing sketching algorithms. 
We need algorithm designers to step in and come up with improved or new
sketches. Meanwhile, the theory community has already developed a rich pool
of sketching tools that may be relevant to the operator's needs. The
challenge lies in how to effectively collect and formulate these requirements  to
motivate algorithm designers to develop new algorithms or disprove the
feasibility.

\subsection{Algorithm Designer \& Platform Vendor/Developer}\label{sec:alg_plat}

\begin{insight}[Sketch Implementation] 
Given a sketch configuration $s\in \Sk$ with device $d$, 
generate a sketch implementation $\csd$ to minimize the
actual resource usages $\rsd$ and $\lsd$
\end{insight}

Ideally, we want to generate optimized platform-specific sketch  implementations for any sketch configuration and device instance. Today, this  requires significant  effort from
both platform vendors/developers and algorithm designers to deliver
 optimized sketch implementation per
hardware target~\cite{SketchVisor,univmon,HashPipe,yang2020joltik}.  What is missing today are tools (e.g., optimizing compilers) to take as input an algorithm definition and configuration defined
in a high-level language, and automatically output an implementation
that is optimized for a particular hardware target. With such a tool,
algorithm designers will not need to worry about how to implement a current
or future sketching algorithm into the hardware architecture and platform developers will not worry about understanding the algorithmic details in order to implement the sketches. 
Existing efforts on P4 language and its target-specific compilers are
expected to contributing in this direction. Unfortunately, our benchmark demonstrates that existing sketch implementations on programmable switches using P4 are far from resource-efficient (Table~\ref{tab:hwresource})\footnote{Sketch configurations in the table, R:rows, C:columns, and L:levels. CountSketch(R=5, C=2048), UnivMon(L=16, R=5, C=2048), R-HHH(L=25, R=5, C=2048), SketchLearn(L=112, R=1, C=2048).}. Compared to a fully functional switch implementation (switch.p4), existing sketches use excessive switch hardware resources (e.g., up to $15\times$ more hash function calls and $17\times$ more stateful ALUs).

\begin{table}[t]
\centering
\scriptsize{
\begin{tabular}{|l|c|c|c|c|}
\hline
{\bf Resource}            &CountSketch&UnivMon& R-HHH & SketchLearn    \\ \hline
Match Crossbar    & 10.0\%       & 177.2\%  & 476.9\%  & 347.7\%\\
SRAM     & 3.5\%        & 56.3\%   & 88.4\%   & 78.9\%\\
Hash Bits   & 3.7\%        & 59.6\%   & 91.6\%   & 82.1\%\\
Hash Func. Calls & 62.5\%       & 1100.0\% & 1562.5\% & 700.0\%\\
Stateful ALUs   & 71.4\%       & 1142.9\% & 1785.7\% & 1600.0\%\\ \hline
\end{tabular}}
\vspace{-2mm}
\caption{ Additional H/W resource usage in Barefoot Tofino by existing sketch implementations. The numbers are normalized by the usage of baseline switch.p4}
\vspace{-3mm}
\label{tab:hwresource}
\end{table}

Recent efforts have focused on performance bottlenecks of sketching
algorithms run inside virtual software switches~\cite{SketchVisor, liu2019nitrosketch,elasticsketch}. While they address the compute/memory bottlenecks in various software sketch implementations,  their ideas
do not directly transfer to other hardware platforms. For instance,
NitroSketch~\cite{liu2019nitrosketch} increases the memory footprint
to reduce CPU consumption, but the key resource constraints in hardware context
are different (e.g., processing stages, ALU, and hash function calls)~\cite{tofino}. SketchVisor~\cite{SketchVisor} and 
ElasticSketch~\cite{elasticsketch} split a sketch into a fast path and
a slow path, and use the fast path to accelerate the packet processing.
This type of idea is not particularly useful in hardware switches where
all packet operations should stay in the fast path~\cite{rmt}.

\begin{mycorollary}[Multi-Sketch Implementation]
 Given all sketch configurations $s \in \Sk$ and device instance $d$,  generate a consolidated sketch implementation $\cd$ for device $d$ such that the actual device resource usage $\rd$ and performance latency $\ld$ are minimized?
\end{mycorollary}

This extension  is about optimizing the sketch implementation on a device when multiple sketch instances are present. Our observation is that many sketches share common primitive operations (hash computation, counter updates, etc.), and we expect that the actual resource usage and packet processing performance on a device $d^*$ can be further optimized to less than $\sum_s r_{s,d^*}$ and $\sum_s l_{s,d^*}$.

A recent proposal~\cite{gao2019autogenerating} shows the promises of using program synthesis to auto-generate fast processing hardware implementations on programmable switches using fewer hardware resources. While this direction is promising  in general for Problem 4 and its extension, this work is a preliminary demonstration in one particular hardware architecture and we would like to see if a similar approach can be designed for other platforms and how many more resources it can save.

\subsection{Network Operator \& Platform Vendor/Developer}

\begin{insight}[Sketch Configuration]
Given a set of traffic workload characteristics $\W$ and traffic-oblivious sketch configurations $S$ that meet accuracy requirements $\Req$,  output a minimal platform-agnostic memory configuration for $\forall s \in \Sk$ that meets the accuracy requirement
\end{insight}

This problem entails finding a minimal memory configuration that meets a certain accuracy requirement for a sketch and a given type of traffic workload characteristics (e.g., skewness, number of flows). Problem 3 attempts to provide a traffic-oblivious memory configuration for the sketch to meet the accuracy requirement under any workloads. For platform vendors, it is of importance to fully understand the resource-accuracy usage of the user functions running atop their platforms and to continue improving cost-efficiency of their architecture. In practice, network operators shall have basic understanding and expectation about the workloads such as skewness and distribution, and the traffic-oblivious configuration may not be tight anymore. For instance, Count Sketch can achieve better memory-accuracy tradeoff if the workload is skewed following some Zipfian distribution~\cite{CountSketch}.

SketchLearn~\cite{sketchlearn} leverages automated statistical inference to actively ``learn'' the traffic workload characteristics to configure its sketch on the fly, relieving the user burdens in the sketch memory configuration. While a learning-based approach is promising in resolving this problem, SketchLearn did not tackle the configurations of other types of sketches and we are unsure whether the model inference used in SketchLearn is an optimal choice.

\begin{insight}[Verification]
Given sketch implementation $\cd$ on device $d$,  ensure that $\cd$ will correctly meet the accuracy requirements when running on $d$?
\end{insight}

Once sketch implementations have been deployed to various devices, one question is that whether the on-device sketch instances will work as expected. Specifically, when an adversary is present, network operators want to verify the integrity of the sketch instances such that the output is correctly reflecting the network traffic conditions. We can think of this verification in two aspects: (1) Network operators can naturally verify the accuracy of sketches if the integrity of the on-device sketch instance is guaranteed.
(2) If such integrity cannot be guaranteed, operators need to identify the occurrences when sketches failed to meet the accuracy requirements. 
Current platform vendors have been on an active race to offer secure enclave primitives on various hardware targets such as Intel SGX, AMD SEV, and ARM TrustZone for mapping arbitrary functions to trusted memory. It remains an open challenge on how to leverage secure hardware capabilities as the ``root-of-the-trust'' for sketch-based telemetry.

Existing efforts \cite{han2017sgx-box, poddar2018safebricks, trach2018shieldbox} demonstrate the promises of protecting network functions with hardware enclaves (e.g., Intel SGX). However, those efforts are not capable of sketch-based telemetry because (1) sketches require high throughput guarantees while existing frameworks such as SafeBricks~\cite{poddar2018safebricks} and SGX-Box~\cite{han2017sgx-box} incur high processing overhead, and (2) these efforts are designed for general-purpose network functions where redundant modules and complexities are included. 

\section{A Future Roadmap}

\begin{figure}[t]
\centering
\includegraphics[width=\linewidth]{figures/vision.pdf}
\vspace{-6mm}
\tightcaption{Sketch-based telemetry framework and stakeholder interactions.}

\label{fig:vision}
\end{figure}

We envision a sketch-based telemetry framework as depicted in 
Figure~\ref{fig:vision}, assuming that research challenges P1-P6 
described above and others have been properly addressed by the 
community. In this framework, we expect a {\em management interface} 
that has an expressive front-end/API to interact with network 
operators, algorithm designers, and platform vendors/developers.
Some key components in the interface are 1) query compiler to 
translate operator intents into sketch configurations, and 2) 
sketch library to maintain state-of-the-art sketching 
definitions/implementations. In the {\em control plane}, there
will be a network-wide resource manager taking input from the management
interface and computing an optimized sketch placement and resource allocation based on the requirements. In the {\em data plane}, the optimized and verified 
sketch instances will be initialized across a network of heterogeneous
devices based on the resource management decisions from the control plane.

We expect 
network operators, algorithm designers, and platform vendors/developers will have a way interacting with the telemetry framework as follows: 

\begin{packeditemize}
\item {\bf Network operators:} Operators can specify their telemetry needs 
via the management interface and receive the intended telemetry metrics via API. In the back-end, operator queries are translated to sketch configurations and their related device-level implementations to be deployed. In addition, operators can also describe their intents to cover some unsupported telemetry tasks.

\item {\bf Algorithm designers:} Algorithm designers can obtain new telemetry capability requests from operators and design new algorithms based on the requests. They can then add their new algorithms to the sketch ecosystem and get feedback about their implemented and evaluated algorithms in real-world scenarios.

\item {\bf Platform vendors and developers:} 
Platform vendors can receive new hardware capabilities requests, deliver new hardware capabilities, and update the device specifications accordingly. Platform developers can explore the sketch algorithm definitions and hardware capabilities in the sketching ecosystem, and deliver improved or new implementations to the sketch library.

\end{packeditemize}

Prior efforts have laid the  groundwork for designing  sketches and making them transition from a theoretical curiosity to a  promising start for network telemetry.  We hope  that  our vision, research challenges,  and collaborative   efforts from  the stakeholders taken together can help transition  sketch-based telemetry into ``prime time'' deployment.

\bibliographystyle{abbrv} 
\begin{small}
\bibliography{alan}
\end{small}

\end{document}